\documentclass[aps,nofootinbib,superscriptaddress]{revtex4}
\usepackage{amstext,amsmath,amssymb,amsfonts,bbm}
\usepackage[latin1]{inputenc}
\usepackage{fancyhdr}
\usepackage[dvips]{graphicx}
\usepackage{epsfig}

\def\f{\frac}
\newcommand{\SU}{\mathrm{SU}}

\def\tl{\widetilde}
\def\pp{\partial}
 
\def\eps{\epsilon}

\def\hhat{\widehat}
\newcommand{\di}[1]{d_{j_{#1}}}
\def\thet{\vartheta}
\newcommand{\curly}[1]{\mathcal{#1}}

\def\be{\begin{equation}}
\def\ee{\end{equation}}
\def\bes{\begin{eqnarray}}
\def\ees{\end{eqnarray}}

\newcommand{\Ref}[1]{(\ref{#1})}

\begin{document}
\title{\large \bf Towards the graviton from spinfoams:\\ the complete perturbative expansion of the 3d toy model}

\author{Valentin Bonzom}\email{valentin.bonzom@ens-lyon.fr}
\affiliation{Centre de Physique Th\'eorique, CNRS-UMR 6207,  Luminy Case 907, 13007 Marseille, France EU}
\affiliation{Laboratoire de Physique, ENS Lyon, CNRS-UMR 5672, 46 All\'ee d'Italie, 69007 Lyon, France EU}

\author{Etera R. Livine}\email{etera.livine@ens-lyon.fr}
\affiliation{Laboratoire de Physique, ENS Lyon, CNRS-UMR 5672, 46 All\'ee d'Italie, 69007 Lyon, France EU}

\author{Matteo Smerlak}\email{matteo.smerlak@ens-lyon.fr}
\affiliation{Laboratoire de Physique, ENS Lyon, CNRS-UMR 5672, 46 All\'ee d'Italie, 69007 Lyon, France EU}

\author{Simone Speziale}\email{sspeziale@perimeterinstitute.ca}
\affiliation{Perimeter Institute, 31 Caroline St. N., Waterloo, ON N2L 2Y5, Canada}

\date{\small February 27, 2008}

\begin{abstract}
We consider an exact expression for the 6j-symbol for the isosceles tetrahedron, involving $\SU(2)$
group integrals, and use it to write the two-point function of 3d gravity on a single tetrahedron
as a group integral. The perturbative expansion of this expression can then be performed with
respect to the geometry of the boundary using a simple saddle-point analysis. We derive the
complete expansion in inverse powers of the length scale and evaluate explicitly the quantum
corrections up to second order. Finally, we use the same method to provide the complete expansion
of the isosceles 6j-symbol with the explicit phases at all orders and the next-to-leading
correction to the Ponzano-Regge asymptotics.
\end{abstract}

\maketitle

\section*{Introduction}

A wide-spread expectation from a full theory of quantum gravity is the possibility to fix the
coefficients appearing in the conventional non-renormalizable perturbative expansion seen as an
effective field theory (EFT). To address this question, a necessary tool is to control the
perturbative expansion of the full theory.
%
%
In this paper, we investigate this issue in the spinfoam formalism, using the 3d toy model with a
single dynamical variable introduced in \cite{3d toy model} and developed in \cite{corrections to
3d model}.

Pursuing a matching with the EFT, while right at the root of many approaches to quantum gravity,
most notably string theory and the asymptotic safety scenario, has long been  obstructed in the
spinfoam formalism. This is due to the difficulty in consistently inserting a background metric to
perform the perturbative expansion. The key idea is to relate the $n$-point functions to the field
propagation kernel, via the introduction of a suitable boundary state \cite{Modesto}.
%
%
The boundary state can then be taken to be a coherent state\footnote{Similar ideas on the use of
coherent states lie also behind the study the semiclassical limit in the canonical loop gravity
framework \cite{CS}.} peaked on a classical geometry \cite{Bianchi}.
We then expect the boundary geometry to effectively induce a semi-classical background structure in
the bulk, which allows to define the graviton propagator from  background-independent correlation
functions.



The structure of this framework is particularly clear in 3d. Considering for simplicity the
Riemannian case, the spinfoam amplitude for a single tetrahedron is the 6j-symbol of the
Ponzano-Regge model. Its large spin asymptotics is dominated by exponentials of the Regge action
for 3d general relativity. This is a key result, since the quantization of the Regge action is
known to reproduce the correct free graviton propagator around flat spacetime \cite{Ruth}. The role
of the boundary state is to induce the flat background and to gauge-fix the propagator
\cite{Reggae}. Thus the framework provides a clear bridge to Regge calculus as an effective
description of spinfoam gravity. However, there is more to it. Indeed, if one works with quantum
Regge calculus alone, there are technical problems to go beyond the free theory approximation.
These are related to the lack of a unique measure for the path integral compatible with the
triangle inequalities conditions ensuring that the metric is positive definite. The issue is solved
in the spinfoam formalism, where the triangle conditions are automatically imposed on the 6j-symbol
by the recoupling theory of $\SU(2)$ and the measure is selected by the topological symmetry of the
system. Thus the spinfoam approach does reduce to quantum Regge calculus at leading order but
improves it beyond \footnotemark.
\footnotetext{The situation is more complicated in 4d.
Developments of this idea have led to the remarkable result that the Barrett-Crane model in 4d
Riemannian spacetime does reproduce at large scales the scaling behavior of the free graviton
propagator (or 2-point function) \cite{Rovelli, etera simone, Bianchi2, grav, Alesci}. This is
crucial evidence towards the correctness of the semiclassical limit of LQG. However the same
developments also pointed out \cite{Bianchi, Alesci} that the Barrett-Crane model does not
reproduce the right tensorial structure of the propagator, thus the model fails to reproduce
General Relativity in the large scale limit. These results have confirmed the validity of the
method, and spurred new efforts towards a better understanding of the spinfoam dynamics
\cite{newvertices,newverticesEPR}. This better behaved models should have a semiclassical limit given by a
modified Regge calculus where the fundamental variables are area and angles, as the one
investigated in \cite{Reggae2}.}

In this paper we consider the simplest possible setting given by the 3d toy model introduced in
\cite{3d toy model, corrections to 3d model} and study analytically the full perturbative
expansion of the 3d graviton. Our results are based on a reformulation of the 6j-symbol and the
graviton propagator as group integrals
and the saddle point analysis of these integrals. We compute explicitly the leading order then both
next-to-leading and next-to-next analytically and we support these results with numerical data.
Moreover, it was shown in \cite{corrections to 3d model} that deviations of the 6j-symbol  from the
leading order Ponzano-Regge asymptotics do not contribute to the next-to-leading order of the
graviton, but that they enter the next-to-next order corrections. Here, the exact representation of
the graviton propagator as a group integral naturally incorporates these deviations.
Finally, an interesting side-product of our calculations is a formula for the next-to-leading order
of the famous Ponzano-Regge asymptotics of the 6j-symbol in the special isosceles configuration.


In spite of the simplicity of the model, the framework we develop here has rather generic features
useful for computing graviton correlation functions in non-perturbative quantum gravity from
spinfoam amplitudes, although it does not allow us to tackle the more general issue of the
existence of a relevant boundary state and of the resulting EFT-like expansion of the correlation
functions for a generic spinfoam triangulation.
We nevertheless show that the full perturbative expansion of the two-point function in the spinfoam
quantization of 3d gravity is computable.
We hope to apply these same methods and tools to 4d spinfoam models and allow a more thorough study
of the full non-perturbative spinfoam graviton propagator and correlations in 4d quantum gravity.




\section{The kernel and the propagator as group integrals}
\subsection{The boundary states and the kernel}

\begin{figure}[ht]
\begin{center}
\includegraphics[width=3cm]{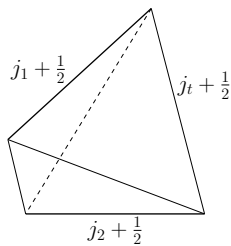}
\end{center}\caption{ \label{tetrahedron setting}
Physical setting to compute the 2-point function. The two edges whose correlations of length
fluctuations will be computed are in fat lines, and have length $j_1+\f{1}{2}$ and
$j_2+\f{1}{2}$. These data are encoded in the boundary state of the tetrahedron. In the time-gauge
setting, the four bulk edges have imposed lengths $j_t+\f{1}{2}$ interpreted as the proper time of
a particle propagating along one of these edges. Equivalently, the time between two planes
containing $e_1$ and $e_2$ has been measured to be $T=(j_t+\f{1}{2})/\sqrt{2}$.}
\end{figure}

Let us consider a triangulation consisting of a single tetrahedron. To define transition amplitudes
in a background independent context for a certain region of spacetime, the main idea is to perform
a perturbative expansion with respect to the geometry of the boundary. This classical geometry acts
as a background for the perturbative expansion. To do so we have to specify the values of the
intrinsic and extrinsic curvatures of such a boundary, that is the edge lengths and the dihedral
angles for a single tetrahedron in spinfoam variables. Following the framework set in \cite{3d toy
model}, we restrict attention to a situation in which the lengths of four edges have been measured,
so that their values are fixed, say to a unique value $j_t+\f{1}{2}$. These constitute the
time-like boundary and we are then interested in the correlations of length fluctuations between
the two remaining and opposite edges which are the initial and final spatial slices (see figure
\ref{tetrahedron setting}). This setting is referred to as the time-gauge setting. The two opposite
edges $e_1$ and $e_2$ have respectively lengths $j_1+\f{1}{2}$ and $j_2+\f{1}{2}$. In the spinfoam
formalism, and in agreement with 3d LQG, lengths are quantized so that $j_t$, $j_1$ and $j_2$ are
half-integers.

The lengths and the dihedral angles are conjugated variables with regards to the boundary geometry,
and have to satisfy the classical equations of motion. Here, it simply means that they must have
admissible values to form a genuine flat tetrahedron. Note that the
dimension of the $\SU(2)$-representation of spin $j$, $d_j\equiv 2j+1$ is twice the edge length.
Setting $k_e=\f{d_{j_e}}{2d_{j_t}}$, for
$e=e_1,e_2$, the dihedral angles $\thet_1$, $\thet_2$ and $\thet_t$ can be expressed in terms of
the lengths :
\be
\thet_{1,2}=2 \arccos \Big(\f{k_{2,1}}{\sqrt{1-k_{1,2}^2}} \Big) \quad\mathrm{and}\quad \thet_t=\arccos\Big( \f{-k_1 k_2}{\sqrt{1-k_1^2}\sqrt{1-k_2^2}} \Big)
\ee
provided $k_e<1$, a condition ensured by the triangle inequalities. Notice the relation : $\cos\
\thet_t= -\cos(\f{\thet_1}{2})\cos(\f{\thet_2}{2})$.

We then need to assign a quantum state to the boundary, peaked on the classical geometry of the
tetrahedron. Since $j_t$ is fixed, we only need such a state for $e_1$, peaked on the length
$j_1+\f{1}{2}$, and for $e_2$, peaked on $j_2+\f{1}{2}$. The previous works used a Gaussian ansatz
for such states. However, it is more convenient to choose states which admit a well-defined Fourier
transform on $\SU(2)$. In this perspective, the dihedral angles of the tetrahedron are interpreted
as the class angles of $\SU(2)$ elements. As proposed in
\cite{corrections to 3d model}, the Gaussian ansatz can be replaced for the edges $e_1$ and $e_2$
by the following Bessel state:
\begin{gather} \label{bound state bessel}
\Psi_e(j) = \f{e^{-\gamma_e/2}}{N}\ \Big[ I_{\lvert j-j_e\rvert}(\f{\gamma_e}{2}) - I_{j+j_e+1}(\f{\gamma_e}{2})\Big]\ \cos(d_j\alpha_e) \\
\text{with}\qquad  \gamma_e=\di{t}(1-k_e^2)
\end{gather}
where $N$ is a normalization coefficient depending on $\gamma_e$. The functions $I_n(z)$ are
modified Bessel functions of the first kind, defined by : $I_n(z) = \f{1}{\pi}\int_0^\pi d\phi\
e^{z\cos\phi}\cos(n\phi)$, and $\alpha_e=\thet_e/2$ is half the dihedral angle. The asymptotics
reproduce the Gaussian behavior peaked around $j_e$, with $\gamma_e$ as the squared width:
\be \label{asymptotics bound state}
\Psi_e(j) = \f{1}{N}\sqrt{\f{4}{\pi\gamma_e}}\ e^{-\f{(j-j_e)^2}{\gamma_e}}\ \cos(d_j\alpha_e).
\ee
The role of the cosine in \eqref{bound state bessel} is to peak the variable dual to $j$, i.e. the
dihedral angle, on the value  $\alpha_e$. Then the boundary state admits a well-defined Fourier
transform, which is a Gaussian on the group $\SU(2)$. We parameterize $\SU(2)$ group elements as
$$
g(\phi, \hat n) = \cos\phi \, \mathbbm{1}+ i \, \sin\phi \, \hat n\cdot
\vec\sigma,
\quad \phi\in[0,2\pi[, \quad \hat{n}\in{\cal S}^2,
$$
where $\sigma_i$ are the Pauli matrices, satisfying $\sigma_i^2=\mathbbm{1}$, and $\phi$ is the
class angle of $g$. Since the group element $g(\phi, \hat n)$ is  identified to $g(\phi+\pi, -\hat
n)$, we can restrict $\phi$ to live in $[0,\pi)$. The Fourier transform of
\eqref{bound state bessel} is then given by:
\begin{gather} \label{boundary state}
\widehat{\Psi}_e (\phi) = \f{1}{2}\ \sum_{\eta=\pm 1} \widehat{\Psi}_e^{(\eta)}(\phi) \\
\text{with}\quad\ \widehat{\Psi}_e^{(\eta)} = \f{1}{N\sin(\phi)}\ \sin\big(d_{j_e}(\phi+\eta\ \alpha_e)\big)\ e^{-\gamma_e\sin^2(\phi+\eta\ \alpha_e)}
\end{gather}
This state is a class function on $\SU(2)$, but $\widehat{\Psi}^{(\eta)}$ alone is not (due to the
$\phi\leftrightarrow-\phi$ symmetry reflecting that a $\SU(2)$ group element and its inverse are
simply related by conjugation). The semiclassical analysis is crystal-clear : it is peaked around
the angle $\alpha_e$ or $\pi-\alpha_e$, according to the sign of $\eta$. The sine shifts the mean
length to $j_e+\f{1}{2}$.

These states carry the information about the boundary geometry necessary to induce a perturbative
expansion around it. More precisely, we are interested in the following correlator,
\begin{align} \label{exact propagator}
&W_{1122}=\f{1}{\curly{N}}\sum_{j'_1,j'_2}\ \begin{Bmatrix} j'_1 & j_t & j_t \\ j'_2 & j_t & j_t \end{Bmatrix}\ \curly{O}_{j_1}(j'_1) \Psi_{e_1}(j'_1)\ \curly{O}_{j_2}(j'_2) \Psi_{e_2}(j'_2) \\
&\text{with}\quad\ \curly{O}_{j_e}(j')=\f{1}{\di{e}^2}\Big(d_{j'}^2-\di{e}^2\Big)
\end{align}
where the normalisation factor $\curly{N}$ is given by the same sum, without the observable
insertions $\curly{O}_{j_e}(j'_e)$. $W_{1122}$ measures the correlations between
length fluctuations for the edges $e_1$ and $e_2$ of the tetrahedron, and it can be interpreted as the
2-point function for gravity \cite{3d toy model}, contracted along the directions of $e_1$ and
$e_2$.

The 6j-symbol, as it enters \eqref{exact propagator}, emerges from the usual spinfoam models for 3d
gravity as the amplitude for a single tetrahedron. In the previous work \cite{corrections to 3d
model} we studied the perturbative expansion using its well-known (leading order) asymptotics in
term of the discrete Regge action (for the tetrahedron). Here instead we use the fact that the
6j-symbol for the isosceles configuration admits an exact expression as group
integrals,\footnote{For a general configuration, one has to consider the squared 6j-symbol to have
an integral expression.}
\be \label{6j}
\begin{Bmatrix}
j_1 & j_t & j_t \\
j_2 & j_t & j_t
\end{Bmatrix} = \int_{\SU(2)^2} dg_1 dg_2\ \chi_{j_t}(g_1g_2)\ \chi_{j_t}(g_1g_2^{-1})\ \chi_{j_1}(g_1)\ \chi_{j_2}(g_2)
\ee
where $\chi_j(g)=\f{\sin (d_j\phi_g)}{\sin\phi_g}$ is the $\SU(2)$ character.
Then, selecting a specific boundary state as described below, we are able to rewrite also
\Ref{exact propagator} as an integral over $\SU(2)$. This allows us to study the perturbative expansion as
the saddle point (or stationary phase) approximation of the integral for large lengths, $\di{e}\gg
1$. With respect to \cite{corrections to 3d model}, this procedure has the advantage of including
the higher order corrections coming from both the Regge action and the corrections to the $\{6j\}$
asymptotics. We will come back to this point below.

Let us begin by looking at the saddle points of the isosceles 6j-symbol,
as the computation of the propagator will have a similar structure.
We first need the angle of the group elements $g_1g_2$ and $g_1g_2^{-1}$ :
\be
\phi_{12}^\pm=\arccos\big( \cos\phi_1 \cos\phi_2 \mp u\ \sin\phi_1 \sin\phi_2 \big)
\ee
where we used the notation $u=\vec{n}_1\cdot\vec{n}_2$. Then, expanding the rapidly oscillatory
phases in exponential form yields:
\be \label{6j exp form}
\begin{Bmatrix}
j_1 & j_t & j_t \\
j_2 & j_t & j_t
\end{Bmatrix}
=\f{1}{8\pi^2}\sum_{\eps_1, \eps_2, \eps_{12}^+, \eps_{12}^- =\pm 1} \eps_1\ \eps_2\ \eps_{12}^+\
\eps_{12}^-\ \int d\phi_1 d\phi_2 du\ f(\phi_1,\phi_2,u)\ e^{i\di{t}\Phi_{\{\eps\}}}
\ee
with
\begin{align} \label{measure f}
f(\phi_1,\phi_2,u) &= \f{\sin(\phi_1)\sin(\phi_2)}{\sin(\phi_{12}^+)\sin(\phi_{12}^-)}, \\
\Phi_{\{\eps\}}(\phi_1,\phi_2,u) &= (\eps_{12}^+\phi_{12}^+ + \eps_{12}^-\phi_{12}^-)\ +\ 2k_1\eps_1\ \phi_1\ +\ 2k_2\eps_2\ \phi_2 \label{6j phases}
\end{align}

Let us proceed to the search for the stationary points of the phase $\Phi_{\{\eps\}}$. The variable
$u$ only enters $\phi_{12}^\pm$, and the related equation, $\eps_{12}^+\pp_u \phi_{12}^+ +
\eps_{12}^-\pp_u \phi_{12}^-=0$, is solved by $u=\vec{n}_1\cdot\vec{n}_2=0$ and
$\eps_{12}^+=\eps_{12}^-=\eps_{12}$. The variational equations with respect to $\phi_1$ and
$\phi_2$ are:
\be
d_{j_t}\big(\eps_{12}^+\pp_{\phi_e}\phi_{12}^+ + \eps_{12}^-\pp_{\phi_e}\phi_{12}^-\big)+\di{e}\eps_e=0 \qquad e=1,2
\ee
and are solved in $[0,\pi[$ by :
\be \label{6j stationary points}
\left\{ \begin{array}{l}
\bar{\phi}_1=-\eps_{12}\ \eps_2\arccos \Big(\f{k_2}{\sqrt{1-k_1^2}}\Big)+ (1+\eps_{12}\ \eps_2) \f{\pi}{2} \\
\bar{\phi}_2=-\eps_{12}\ \eps_1\arccos \Big(\f{k_1}{\sqrt{1-k_2^2}}\Big)+ (1+\eps_{12}\ \eps_1) \f{\pi}{2}
\end{array}
\right.
\ee
Notice that this result allows us to give a geometrical interpretation to the class angles entering
\Ref{6j exp form}, which is similar to the one for the usual integral formula of the squared 6j-symbol:
the isosceles 6j-symbol is peaked on half the internal, or external, dihedral angles of the
classical geometry. Indeed, for example, when $\eps_{12}\eps_2=-1$, the stationary angle
$\bar{\phi}_1$ is $\bar{\phi}_1=\alpha_1=\thet_1/2$, while for $\eps_{12}\eps_2=1$, we have
$\bar{\phi}_1=\pi-\alpha_1$. \footnote{This geometric interpretation of the saddle point is not
surprising since the 6j-symbol is indeed the unique physical quantum state for a trivial topology
and a triangulation made of a single tetrahedron \cite{physical bound state}. It satisfies the
quantum flatness constraint and can serve as the boundary state in the general boundary framework.}

We perform the complete expansion of the isosceles $\{6j\}$, using this stationary phase analysis,
below in section \ref{isosceles 6j expansion}. Now we turn to the graviton propagator.

\subsection{The propagator as group integrals}

If we insert the expression \eqref{6j} into \eqref{exact propagator}, the sums over $j'_1$ and
$j'_2$ give the SU(2) Fourier transform of the boundary states. Let us first look at the
normalization $\curly{N}$ :
\begin{align}
\curly{N} &= \int dg_1 dg_2\ \chi_{j_t}\big(g_1 g_2^{-1}\big)\ \chi_{j_t}\big(g_1 g_2\big)\ \Big[\sum_{j'_1}\Psi_{e_1}(j'_1)\chi_{j'_1}(g_1)\Big]\ \Big[\sum_{j'_2}\Psi_{e_2}(j'_2)\chi_{j'_2}(g_2)\Big] \\
 &= \int dg_1 dg_2\ \chi_{j_t}\big(g_1 g_2^{-1}\big)\ \chi_{j_t}\big(g_1 g_2\big)\ \widehat{\Psi}_{e_1} (g_1)\ \widehat{\Psi}_{e_2} (g_2) \label{norm group integral}
\end{align}
For large spins, we are interested in evaluating $\curly{N}$ with a saddle point approximation.
To that end, let us expand the previous expression in exponential form :
\begin{align} \label{normalisation exp form}
\curly{N}=&\f{1}{32\pi^2} \sum_{\substack{\eps_1, \eps_2, \eps_{12}^+, \eps_{12}^-,\\ \eta_1, \eta_2 =\pm 1}} \eps_1\ \eps_2\ \eps_{12}^+\ \eps_{12}^-\ \int d\phi_1 d\phi_2 du\ f(\phi_1,\phi_2,u)\ e^{\di{t} S_{\{\eps,\eta\}}(\phi_1,\phi_2,u)} \\
\text{with}\qquad
&S_{\{\eps,\eta\}}(\phi_1,\phi_2,u) = i\big(\eps_{12}^+\phi_{12}^+ + \eps_{12}^-\phi_{12}^-\big) + \sum_{e=1,2} 2ik_e\eps_e(\phi_e+\eta_e\alpha_e) - \big(1-k_e^2\big)\sin^2(\phi_e+\eta_e\alpha_e) \label{action}
\end{align}
where the label $\{\eps,\eta\}$ refers to the dependence on the sign variables, and $f$ is given by
\eqref{measure f}. The crucial point is that the phase of \eqref{normalisation exp form} (the
imaginary part of $S_{\{\eps,\eta\}}$) is precisely $\Phi_{\{\eps\}}$ in \eqref{6j phases}, up to
constant $\alpha_{1,2}$ terms which play no role in the stationary phase approximation. This means
in particular that the imaginary part of $S_{\{\eps,\eta\}}$ has the same saddle points of the
isosceles 6j-symbol.

The same analysis can be performed for the numerator of $W_{1122}$. To take into account the
observables $\curly{O}_{j_e}$, notice that the $\SU(2)$ character is an eigenfunction for the
Laplacian on the sphere $S^3$ with the Casimir as eigenvalue:
\be
\Delta_{S^3}\chi_j(\phi)=\f{1}{\sin^2\phi}\ \pp_\phi\big(\sin^2\phi\ \pp_\phi\ \chi_j\big)=-(d_j^2-1)\ \chi_j(\phi).
\ee
This allows to perform the sums over $j'_1$ and $j'_2$ in \eqref{exact propagator}, introducing the
Fourier transforms $\widehat{\Psi}_{e}$. Again expanding the result of these operations into
exponential form, one ends up with:
\be \begin{split} \label{prop exp form}
W_{1122} = \f{1}{32\pi^2\ \curly{N}}\ \f{k_1 k_2}{4\cos^2\thet_t} \sum_{\substack{\eps_1, \eps_2, \eps_{12}^+, \eps_{12}^-,\\ \eta_1, \eta_2 =\pm 1}} \eps_1\ \eps_2\ \eps_{12}^+\ \eps_{12}^-\ \int d\phi_1 &d\phi_2 du\ f(\phi_1,\phi_2,u)\ e^{\di{t} S_{\{\eps,\eta\}}(\phi_1,\phi_2,u)} \\
&\times\Big( a_{\{\eps,\eta\}}(\phi_1,\phi_2)+\f{b_{\{\eps,\eta\}}(\phi_1,\phi_2)}{\di{t}}+\f{c_{\{\eps,\eta\}}(\phi_1,\phi_2)}{\di{t}^2}\Big)
\end{split} \ee
The functions $a_{\{\eps,\eta\}}$, $b_{\{\eps,\eta\}}$ and $c_{\{\eps,\eta\}}$ stand for the observable insertions, and are given by :
\begin{align}
a_{\{\eps,\eta\}}(\phi_1,\phi_2) &= \prod_{e=1,2} \Big( \f{1-k_e^2}{2k_e}\sin^2 2(\phi_e+\eta_e\ \alpha_e) - 2i\eps_e \sin2(\phi_e+\eta_e\ \alpha_e)\Big) \\
b_{\{\eps,\eta\}}(\phi_1,\phi_2) &= -\f{1}{k_2}\cos 2(\phi_2+\eta_2\ \alpha_2)\ \Big( \f{1-k_1^2}{2k_1}\sin^2 2(\phi_1+\eta_1\ \alpha_1) - 2i\eps_1 \sin2(\phi_1+\eta_1\ \alpha_1)\Big) + \big( e_1\leftrightarrow e_2\big) \\
c_{\{\eps,\eta\}}(\phi_1,\phi_2) &= \f{1}{k_1 k_2}\ \cos 2(\phi_1+\eta_1\ \alpha_1)\ \cos 2(\phi_2+\eta_2\ \alpha_2)
\end{align}

We are now ready to study the large spin expansion of \Ref{prop exp form}. A common choice in the
literature is to do so using a power series in $1/j_t$ (keeping $j_1/j_t$ and $j_2/j_t$ constant).
However it is more convenient to take as parameter of the expansion the dimension $\di{t}$ (again
keeping $k_1,k_2$ fixed). This is the natural choice, as we compute correlations with respect to
the background geometry with lengths defined by the half-dimensions $\di{t}/2$, $\di{1}/2$ and
$\di{2}/2$. Furthermore, as we show below (see \eqref{regge action} and \eqref{tetrahedron
volume}), these are the values of the lengths emerging in the asymptotics of the 6j-symbol.

As written in \eqref{prop exp form}, $W_{1122}$ corresponds to the mean value of the function
$a_{\{\eps,\eta\}}+\f{b_{\{\eps,\eta\}}}{\di{t}}+\f{c_{\{\eps,\eta\}}}{\di{t}^2}$ for the
non-linear theory defined by the action $S_{\{\eps,\eta\}}$ and the integration measure $f$. The
strategy is thus clear: we will compute separately the normalisation $\curly{N}$ and the numerator,
perturbatively, with an expansion around the saddle points of the action $S_{\{\eps,\eta\}}$.


As stated above, the imaginary part of $S_{\{\eps,\eta\}}$ has the same saddle points of the
isosceles 6j-symbol, namely $u=0$ and $\bar{\phi}_e$ given in \eqref{6j stationary points}
(independently of $\eta_1$ and $\eta_2$). The extremization with respect to the real part of
$S_{\{\eps,\eta\}}$, on the other hand, constrains the $\eta_1$ and $\eta_2$ signs. Indeed, for a
given solution $(\bar{\phi}_1,\bar{\phi}_2)$ from \eqref{6j stationary points}, characterized by
$\eps_{12}$, $\eps_1$ and $\eps_2$, the signs $\eta_1$ and $\eta_2$ have to satisfy:
\be
\sin 2\big(\bar{\phi}_e+\eta_e\ \alpha_e\big) = 0,\qquad\text{for}\ e=1,2
\ee
These equations are solved by taking $\eta_1=\eps_2\eps_{12}$ and $\eta_2=\eps_1\eps_{12}$. This
leads to four possibilities, summarized in the following table,
\begin{displaymath}
\begin{array}{c|c|c}
\; & \eta_1=-1 & \eta_1=1 \\
\hline
\eta_2=-1 & \quad \begin{aligned} &\bar{\phi}_1=\alpha_1,\ \text{and}\ \bar{\phi}_2=\alpha_2, \\ &\eps_1=\eps_2=-\eps_{12} \end{aligned} & \quad \begin{aligned} &\bar{\phi}_1=\pi-\alpha_1,\ \text{and}\ \bar{\phi}_2=\alpha_2, \\ &\eps_1=-\eps_2=-\eps_{12} \end{aligned}  \\
\hline
\eta_2=1 & \quad \begin{aligned} &\bar{\phi}_1=\alpha_1,\ \text{and}\ \bar{\phi}_2=\pi-\alpha_2, \\ &-\eps_1=\eps_2=-\eps_{12} \end{aligned} &\quad \begin{aligned} &\bar{\phi}_1=\pi-\alpha_1,\ \text{and}\ \bar{\phi}_2=\pi-\alpha_2, \\ &\eps_1=\eps_2=\eps_{12} \end{aligned}
\end{array}
\end{displaymath}

The condition $\eps_{12}^+=\eps_{12}^-=\eps_{12}$ and $\eta_1=\eps_2\eps_{12}$ and $\eta_2=\eps_1\eps_{12}$
allows us to perform three sums in \Ref{prop exp form},
the configurations for which there is no saddle point being exponentially suppressed :
\be \begin{split}
W_{1122} = \f{1}{32\pi^2\ \curly{N}}\ \f{k_1 k_2}{4\cos^2\thet_t} \sum_{\eps_1, \eps_2, \eps_{12}=\pm 1} \eps_1\ \eps_2\ \int d\phi_1 &d\phi_2 du\ f(\phi_1,\phi_2,u)\ e^{\di{t} S_{\{\eps\}}(\phi_1,\phi_2,u)} \\
&\times\Big( a_{\{\eps\}}(\phi_1,\phi_2)+\f{b_{\{\eps\}}(\phi_1,\phi_2)}{\di{t}}+\f{c_{\{\eps\}}(\phi_1,\phi_2)}{\di{t}^2}\Big)
\end{split} \ee
and the same for $\curly{N}$ without the insertion of $\f{k_1 k_2}{4\cos^2\thet_t}(a_{\{\eps\}}+b_{\{\eps\}}/\di{t}+c_{\{\eps\}}/\di{t}^2)$.
Here the label $\{\eps\}$ simply indicates the dependence of the functions on the signs $\eps_1$, $\eps_2$ and $\eps_{12}$.

\section{The complete perturbative expansion}

The perturbative expansion of the two-point function $W$ is formulated as an asymptotic power
series expansion in $1/\di{t}$, of the type:
\be
W\,=\, \f1{\di{t}} \,\left[w_0+\f1{\di{t}}w_1+\f1{\di{t}^2}w_2+\dots\right].
\ee
Let us remind the reader that the dimension $\di{t}$ defines the length scale of the tetrahedron
$L=\di{t}L_P/2$, with the Planck length $L_P=\hbar G$. Such an expansion thus matches the typical
expansion of quantum field theory correlations with quantum corrections ordered in increasing
powers of $\hbar$ (and of the coupling constant $G$) and with $L$ corresponding to the
renormalization scale. We can thus call the coefficients $w_1,w_2,..$ the one-loop and two-loop
(and so on) corrections.

This perturbative expansion is obtained studying the power series expansion in $1/\di{t}$ around
each of the four saddle points of both the denominator $\int f\exp\di{t}S_{\{\eps\}}$ and the
numerator $\int(a_{\{\eps\}}+b_{\{\eps\}}/\di{t}+c_{\{\eps\}}/\di{t}^2)f\exp\di{t}S_{\{\eps\}}$.
More precisely, we expand the action  $S_{\{\eps\}}$ around its saddle point: the evaluation of $S$
at the stationary point gives a numerical factor, there is no linear term obviously, the quadratic
term defines the Hessian matrix $A_{\{\eps\}}$ and finally all the remaining higher order terms
(cubic onwards) are kept together to define the potential $\Omega_{\{\eps\}}$. This potential thus
contains all higher order corrections to the quadratic approximation to the action S. As such, it
does not enter the leading order of the two-point function but largely enters its NLO, NNLO and so
on, (the loop corrections) as in quantum field theory.
Then each term in the power series is evaluated as the Gaussian moment with respect to the Hessian
matrix $A_{\{\eps\}}$ of terms coming from the expansion of $f\exp\di{t}\Omega_{\{\eps\}}$ in
powers of $\di{t}$.
In general many terms actually contribute to the same overall order in $1/\di{t}$. This intricate
structure comes about precisely as in \cite{corrections to 3d model} because the expansion of
$\exp\di{t}\Omega_{\{\eps\}}$ gives increasing powers of $\di{t}$ while the Gaussian moments have
increasing powers in $1/\di{t}$.


On the other hand, a simplification of our calculations comes from the fact that each saddle point
gives the same contribution. This is a consequence of the symmetry properties of the functions
involved under the transformation of $\phi_e$ into $\pi-\phi_e$. Further, for a given saddle point,
the two possible configurations of signs are simply related by complex conjugation. The actual sum
then ensures the reality of the result. Without loss of generality, we can thus restrict the
computation to the saddle point $(\alpha_1,\alpha_2,0)$ with $\eps_{12}=1=-\eps_1=-\eps_2$. There
we have :
\begin{align}\label{a}
a(\phi_1,\phi_2) &= \prod_{e=1,2} \Big( \f{1-k_e^2}{2k_e}\sin^2 2(\phi_e- \alpha_e) + 2i \sin2(\phi_e- \alpha_e)\Big) \\
b(\phi_1,\phi_2) &= -\f{1}{k_2}\cos 2(\phi_2- \alpha_2)\ \Big( \f{1-k_1^2}{2k_1}\sin^2 2(\phi_1- \alpha_1) + 2i \sin2(\phi_1-\alpha_1)\Big) + \big( e_1\leftrightarrow e_2\big) \\
c(\phi_1,\phi_2) &= \f{1}{k_1 k_2}\ \cos 2(\phi_1-\alpha_1)\ \cos 2(\phi_2-\alpha_2)
\end{align}
and the potential $\Omega$ is extracted from the derivatives of $S$ greater than three, with $S$
given by \Ref{action} with the chosen signs,
\be
S(\phi_1,\phi_2,u) = i\big(\phi_{12}^+ + \phi_{12}^-\big) - \sum_{e=1,2} \big(1-k_e^2\big)\sin^2(\phi_e-\alpha_e)+\mathrm{linear\ terms}
\ee
Expanding around the background, the inverse of the Hessian matrix is (see the Appendix A for details):
\be \label{Ainv}
A^{-1} = \f{1}{4} \begin{pmatrix} \f{1}{1-k_1^2} & \f{\cos\thet_t}{k_1 k_2}\ e^{i\thet_t} & 0 \\
                                                        \f{\cos\thet_t}{k_1 k_2}\ e^{i\thet_t} & \f{1}{1-k_2^2} & 0 \\
                                                        0 & 0 & \f{2i\ \tan\thet_t}{1-(k_1^2+k_2^2)} \end{pmatrix}.
\ee
Introducing the shorthand notation
\be
A^{-1}_{\vec{\beta}} = \sum_{\substack{\mathrm{all\ possible\ pairings} \\ \mathrm{of}\ (\beta_1,\dots,\beta_{2N})}} A^{-1}_{\beta_{i_1}\beta_{i_2}}\dots A^{-1}_{\beta_{i_{2N-1}}\beta_{i_{2N}}}
\ee
for $\vec{\beta}\in\{1,2,3\}^{2N}$, the complete perturbative expansion of the propagator can be written as
\be \label{prop final expansion}
W_{1122} \,=\,
\f{k_1 k_2}{4\cos^2\thet_t}\,
\f{\f{\sqrt{1-k_1^2}\sqrt{1-k_2^2}}{2\di{t}}\sum_{i,j=1,2}\pp^2_{ij}a\ A^{-1}_{ij} + \sum_{P\geq2}
\f{W_P}{\di{t}^P}}{\sum_{P\in\mathbb{N}} \f{N_P}{\di{t}^P}}\, .
\ee
In the numerator, the first term gives the leading order contribution in $1/\di{t}$ (see next
section). It comes entirely from the $a$ term in \Ref{a}. In fact, $a$ and $b$ vanish at the saddle
point, and so does the gradient of $a$, so the expansion of $a$, $b$ and $c$ is dominated by the
quadratic term of $a$.

All the higher order corrections have been collected in the summations. The coefficients $N_P$ and
$W_P$ correspond to finite sums:
\begin{gather} \label{coeff norm}
N_P =
\sum_{n=0}^{2P} \sum_{\vec{\beta}\in\{1,2,3\}^{2(P+n)}}
\f{1}{(2(P+n))!\ n!}\, \Re\Big(i\ e^{-i(2d_{j_t}-\f{1}{2})\thet_t}\,
\pp^{2(P+n)}_{\vec{\beta}}\big(f\Omega^n\big)\ A^{-1}_{\vec{\beta}}\Big)_{|\phi_1=\alpha_1,\phi_2=\alpha_2,u=0} \\
\begin{split} \label{coeff prop}
\text{and}\qquad W_P = \sum_{n\geq0}\f{1}{n!}\Big\{ &\sum_{\vec{\beta_a}\in\{1,2,3\}^{2(P+n)}} \f{1}{(2(P+n))!}\ \Re\Big(ie^{-i(2d_{j_t}-\f{1}{2})\thet_t}\ \pp^{2(P+n)}_{\vec{\beta_a}}\big(af\Omega^n\big)\ A^{-1}_{\vec{\beta_a}}\Big) \\
+ &\sum_{\vec{\beta_b}\in\{1,2,3\}^{2(P+n-1)}} \f{1}{(2(P+n-1))!}\ \Re\Big(ie^{-i(2d_{j_t}-\f{1}{2})\thet_t}\ \pp^{2(P+n-1)}_{\vec{\beta_b}}\big(bf\Omega^n\big)\ A^{-1}_{\vec{\beta_b}}\Big) \\
+ &\sum_{\vec{\beta_c}\in\{1,2,3\}^{2(P+n-2)}} \f{1}{(2(P+n-2))!}\
\Re\Big(ie^{-i(2d_{j_t}-\f{1}{2})\thet_t}\ \pp^{2(P+n-2)}_{\vec{\beta_c}}\big(cf\Omega^n\big)\
A^{-1}_{\vec{\beta_c}}\Big) \Big\}_{|\phi_1=\alpha_1,\phi_2=\alpha_2,u=0}
\end{split}
\end{gather}
The three lines of \eqref{coeff prop} are the three separate contributions of the insertions $a$,
$b$ and $c$. The sum over $n$ defining $W_P$ is finite for each of these contributions : $n$ is
bounded by $2P-2$, $2P-3$ and $2P-4$ for $a$, $b$ and $c$ respectively. The derivatives of highest
order of $\Omega$ involved in $W_P$ are respectively the $2P$-th derivatives, the $(2P-1)$-th ones
and the $(2P-2)$-th ones, and for $N_P$, the $2(P+1)$-th derivatives, all corresponding to $n=1$.

The intricacy of the formulas was anticipated at the beginning of the section. However the reader should
be reassured that they are simple, if tedious, algebraic expressions.

The real part $\Re$ in \Ref{coeff norm} is consistent with the reality of the
initial expression \Ref{prop exp form}, and arises from the summation over the $\epsilon$ sign.

\section{The leading order, one-loop and two-loop corrections}

We now use \eqref{coeff norm} and \eqref{coeff prop} to obtain explicitly the first orders of the
expansion. The leading order (LO) and the next to leading order (NLO) have already been obtained in
a quite different way in \cite{corrections to 3d model}. We here recover them quickly. The
computation of the next to next to leading order (NNLO) is then completely new. It is shown in
\cite{corrections to 3d model} that the NNLO needs the corrections to the asymptotics of the
6j-symbol, i.e. to the Ponzano-Regge formula. The success of our method resides in the fact that
such corrections are naturally contained in the exact group integral expression \eqref{6j} of the
kernel.

The LO is obtained evaluating the normalization at the saddle point, $f_0=f(\alpha_1,\alpha_2,0)$,
and the numerator at the non-zero second derivatives of $a$:
\begin{align}
W^{LO}_{1122} &=
\f{k_1 k_2}{4\cos^2\thet_t}\ \f{\f{1}{\di{t}}
\Re\big(ie^{-i(2d_{j_t}-\f{1}{2})\thet_t}\pp^2_{\phi_1,\phi_2}a\ A^{-1}_{12}\big)}{f_0\Re\big(ie^{-i(2d_{j_t}-\f{1}{2})\thet_t}\big)} \\
 &= -\f{1}{\di{t}\cos\thet_t}\ \f{\sin (2\di{t}-\f{3}{2})\thet_t}{\sin (2\di{t}-\f{1}{2})\thet_t} \label{propLO}
\end{align}
This reproduces the expected $1/\di{t}$ scaling behavior of the LO. The difference in the
coefficient with \cite{3d toy model, corrections to 3d model} comes from the different boundary
state used. In particular notice that while $\thet_t(k_1, k_2)$ is a constant, the dependence upon
$\di{t}$ of the second fraction produces spurious oscillations. These can be reabsorbed in the
boundary state, replacing $\sin \di{e}(\phi+\eta\ \alpha_e)$ in $\hhat{\Psi}_e$ with $\sin
\big(\di{e}(\phi+\eta\ \alpha_e) + \eta\  \di{t}\thet_t\big)$. The Fourier transform is then:
\be \label{simplified bound state}
\Psi_e(j) = \f{e^{-\gamma_e/2}}{N}\ \Big[ I_{\lvert j-j_e\rvert}(\f{\gamma_e}{2})\,
\cos(d_j\alpha_e+\di{t}\thet_t) - I_{j+j_e+1}(\f{\gamma_e}{2})\ \cos(d_j\alpha_e-\di{t}\thet_t)\Big]
\ee
This does not affect the asymptotic behavior of $\Psi_e(j)$. With this replacement, we obtain the
same result of \cite{corrections to 3d model} (cf. equation (37)) for the isosceles case,
\be \label{real LO}
W_{1122}^{LO} = -\f{1}{\di{t}\cos\thet_t}\ \f{\sin \f{3}{2}\thet_t}{\sin \f{1}{2}\thet_t}.
\ee
Even if the LO now matches the previous results presented in \cite{corrections to 3d model}, the
higher orders will differ because of the different boundary state used.

For the sake of a simpler presentation, we will report the NLO and the NNLO for the equilateral
tetrahedron, $k_1=k_2=1/2$ and $\thet_t=\arccos-\f{1}{3}$. The general expressions in terms of
$k_1$ and $k_2$ are indeed quite cumbersome. This choice will also facilitate the comparison with
numerical simulations of \eqref{exact propagator}.

\begin{figure}[t] \begin{center}
\includegraphics[width=5cm]{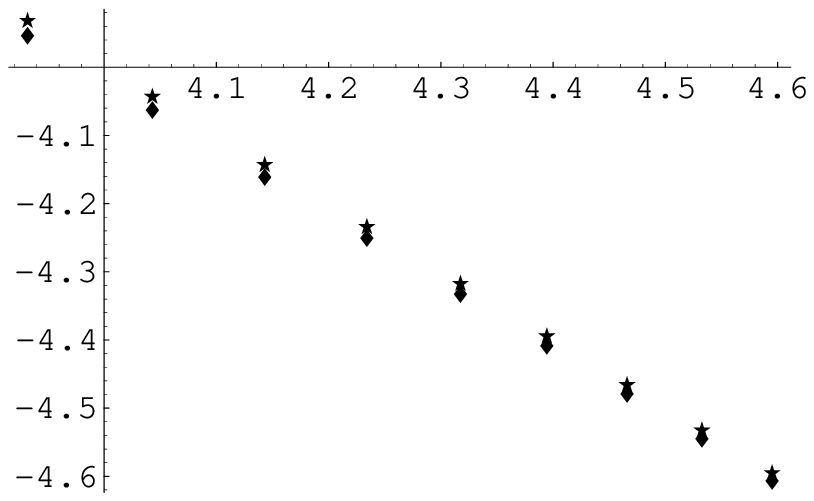}\qquad
\includegraphics[width=5cm]{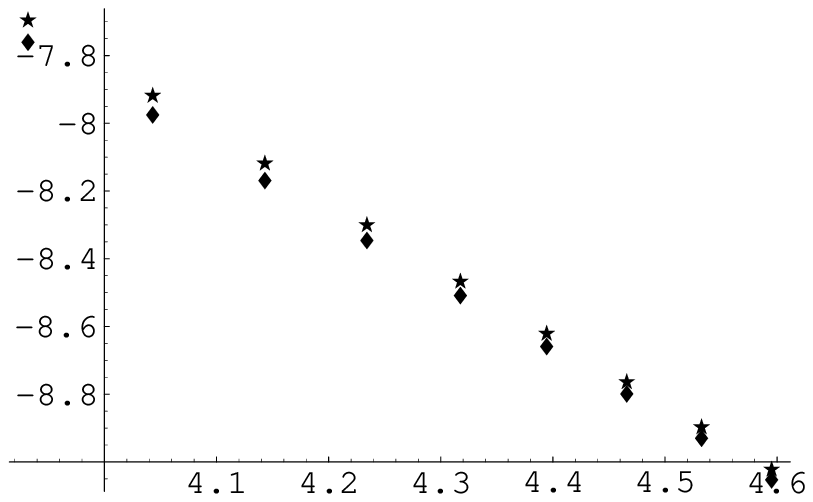}\quad
\includegraphics[width=5cm]{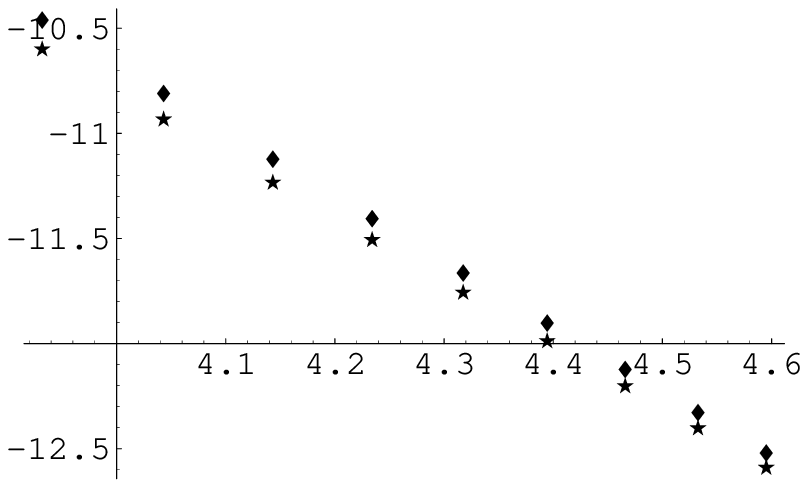}\quad
\caption{ \label{numerics LO NLO NNLO} Log-log plots comparing numerical simulations with analytical results. Left plot: a numerical simulation of \eqref{exact propagator} (diamond symbol) compared with the leading order of \eqref{equilateral NLO} (star symbol). Middle plot: the next to leading order of \eqref{equilateral NLO}, in star shape, with the numerics, in diamond shape. Right plot: the next to next to leading order \eqref{equilateral NNLO}.}
\end{center}
\end{figure}

The NLO is then obtained from the coefficients $N_1$ and $W_2$. To keep compact expressions, we
adopt the following symbolic notation for the contractions of derivatives with Gaussian moments :
for functions $f$ and $h$ of $\phi_1$, $\phi_2$ and $u$, define: $f_n\ h_m\
A^{-1}_{n+m}=\f{1}{n!m!}\sum_{i_1,\cdots,i_n=1,2,3}\sum_{j_1,\cdots,j_m=1,2,3}
\pp^n_{i_1,\cdots,i_n}f\ \pp^m_{j_1,\cdots,j_m}h\ A^{-1}_{(i_1,\cdots,i_n,j_1,\cdots,j_m)}$
evaluated at the saddle point $(\alpha_1,\alpha_2,0)$ with $\eps_{12}=-\eps_1=-\eps_2=1$. For
example, the LO of \eqref{prop final expansion} can be written $\f{1}{\di{t}}a_2 A^{-1}_2$. In
$N_1$, three powers of $\Omega$ appear, $\Omega^0$, $\Omega$ and $\Omega^2$. Using the boundary
state
\eqref{simplified bound state}, we have:
\be
N_1 = \Re\Big(i\ e^{\f{i}{2}\thet_t}\ \Big[f_2\ A^{-1}_2 + \big(f_1\ S_3 + f_0\ S_4\big) A^{-1}_4 + \f{f_0}{2}\ S_3\ S_3\ A^{-1}_6\Big]\Big)
\ee
We proceed in the same way for the three contributions to $W_2$:
\be \begin{split}
W_2 = \Re\Big(i\ e^{\f{i}{2}\thet_t}\ \Big[ \big(a_2\ f_2 + a_3\ f_1 + f_0\ a_4\big)A^{-1}_4 + \big(a_2\ f_1\ S_3 &+ f_0 (a_3\ S_3 + a_2\ S_4)\big)A^{-1}_6 + \f{f_0}{2}\ a_2\  S_3\ S_3\ A^{-1}_8 \\
&+ \big(f_0\ b_2 + b_1\ f_1\big)A^{-1}_2 + f_0\ b_1\ S_3\ A^{-1}_4 + f_0\ c_0\Big]\Big)
\end{split} \ee
After straighforward algebra we obtain the NLO, of order $1/\di{t}^2$:
\be \label{equilateral NLO}
W^{NLO}_{1122} = \f{1}{\di{t}} - \f{511}{432\ \di{t}^2}
\ee
These results for the LO and NLO are well-confirmed by numerical simulations, as one can see from
figure \ref{numerics LO NLO NNLO}. An agreement with 0.58$\%$ of error for the LO, and with 1.7$\%$
error for the NLO is reached between the coefficients of these orders for $\di{t}=201$ (i.e the
representation of spin $j_t=100$).

All orders of the expansion can be computed using the above recipe. From this point of view, the
NNLO is of no particular specificity. We need the expansion of the action (or equivalently
$\Omega$) until the sixth order. The highest order correlator $A^{-1}_{\vec{\beta}}$ is of order 12
for the normalisation, and respectively 14, 10 and 6 for the insertion of $a$, $b$ and $c$.
\be \begin{split}
N_2 = \Re\Big(i\ e^{\f{i}{2}\thet_t}\ \Big[ f_4\ A^{-1}_4 + \big(f_3\ S_3 + f_2\ S_4 &+ f_1\ S_5 + f_0\ S_6\big)A^{-1}_6 + \f{1}{2}\big(f_2\ S_3\ S_3 + 2f_1\ S_3\ S_4 + f_0\ S_4\ S_4 + 2f_0\ S_3\ S_5\big)A^{-1}_8 \\
 &+ \f{1}{3!}\big(f_1\ S_3\ S_3\ S_3 + 3f_0\ S_3\ S_3\ S_4\big)A^{-1}_{10} + \f{f_0}{4!}S_3\ S_3\ S_3\ S_3\ A^{-1}_{12}\Big]\Big)\end{split}
\ee
We also write $W_3 = W_3^{(a)}+W_3^{(b)}+W_3^{(c)}$, with:
\begin{gather} \begin{split}
W_3^{(a)} = \Re\Big(i\ e^{\f{i}{2}\thet_t}\ \Big[&\big(f_0\ a_6 + f_1\ a_5 + f_2\ a_4 + f_3\ a_3 + f_4\ a_2\big)A^{-1}_6 + \big( (f_0\ a_5 + f_1\ a_4 + f_2\ a_3 + f_3\ a_2)S_3 \\
+& (f_0\ a_4 + f_1\ a_3 + f_2\ a_2)S_4 + (f_0\ a_3 + f_1\ a_2)S_5 + f_0\ a_2\ S_6\big)A^{-1}_8 + \f{1}{2}\big( (f_0\ a_4 + f_1\ a_3 + f_2\ a_2)S_3\ S_3 \\
+& 2(f_0\ a_3 + f_1\ a_2)S_3\ S_4 + f_0\ a_2(S_4\ S_4 + 2S_3\ S_5)\big)A^{-1}_{10} + \f{1}{3!}\big((f_0\ a_3 + f_1\ a_2)S_3\ S_3\ S_3 \\
+& 3f_0\ a_2\ S_3\ S_3\ S_4\big)A^{-1}_{12} + \f{f_0}{4!}a_2\ S_3\ S_3\ S_3\ S_3\ A^{-1}_{14}\Big]\Big)
\end{split} \\
\begin{split}
W_3^{(b)} = \Re\Big(i\ e^{\f{i}{2}\thet_t}\ \Big[ \big( &f_0\ b_4 + f_1\ b_3 + f_2\ b_2 + f_3\ b_1\big)A^{-1}_4 + \big( (f_0\ b_3 + f_1\ b_2 + f_2\ b_1)S_3 + (f_0\ b_2 + f_1\ b_1)S_4 \\
+ &f_0\ b_1\ S_5\big)A^{-1}_6 + \f{1}{2}\big( (f_0\ b_2 + f_1\ b_1)S_3\ S_3 + 2f_0\ b_1\ S_3\ S_4\big)A^{-1}_8 + \f{f_0}{3!}b_1\ S_3\ S_3\ S_3\ A^{-1}_{10}\Big]\Big) \end{split} \\
\begin{split}
W_3^{(c)} = \Re\Big(i\ e^{\f{i}{2}\thet_t}\ \Big[ \big(f_0\ c_2 + f_1\ c_1 + f_2\ c_0\big)A^{-1}_2 + \big( (f_0\ c_1 + f_1\ c_0)S_3 + f_0\ c_0\ S_4\big)A^{-1}_4 + \f{f_0}{2}c_0\ S_3\ S_3\ A^{-1}_6\Big]\Big)
\end{split}
\end{gather}
The NNLO is thus computed to be:
\be \label{equilateral NNLO}
W^{NNLO}_{1122} = \f{1}{\di{t}} - \f{511}{432\ \di{t}^2} + \f{520507}{157464\ \di{t}^3}
\ee
This result is again supported by numerical simulations, see figure \ref{numerics LO NLO NNLO}. An
agreement with 11.3$\%$ of error is obtained for the coefficient of $1/\di{t}^3$ at $\di{t}=201$.
The error can be reduced pushing the simulations to higher values of $\di{t}$.

\section{Perturbative expansion of the isosceles 6j-symbol} \label{isosceles 6j expansion}

The procedure described above can be applied directly to the isosceles 6j-symbol \Ref{6j},
obtaining the known Ponzano-Regge formula and its corrections. This is interesting for a number of
reasons. As discussed in \cite{corrections to 3d model}, the corrections to the Ponzano-Regge
formula are a key difference between the spinfoam perturbative expansion studied here, and the one
that would arise from quantum Regge calculus. The 6j-symbol is also the physical boundary state of
3d gravity for a trivial topology and a one-tetrahedron triangulation. In 4d, it appears as a
building block for the spin-foams amplitudes, such as the 15j-symbol. Thus, with regards to many
aspects of spin-foams in 3d and 4d, in particular for the quantum corrections to the semiclassical
limits, it would be good to have a better understanding of this object beyond the Ponzano-Regge
asymptotics. This is what we do in this section, performing a perturbative expansion of the exact
expression
\eqref{6j} for the isosceles 6j-symbol. Indeed, it is a simpler application of the procedure
developed above for the propagator.

As written in \eqref{6j exp form}, the 6j-symbol is the partition function for the theory defined
by the action $\Phi_{\{\eps\}}$ and the integration measure $f$. Notice first of all that the Regge
action of 3d gravity $S_R=\sum \di{e}\f{\thet_e}{2}$ naturally appears as the evaluation of
$\Phi_{\{\eps\}}$ at the saddle points:
\be \label{regge action}
\left\{ \begin{aligned} &e^{i\di{t}\Phi_{\{\eps\}}(\bar{\phi}_1,\bar{\phi}_2,0)} = \eps_1\eps_2\ e^{-i\eps_1\eps_2\eps_{12}S_R} \\
&S_R = \di{t}\big(2\thet_t + 2k_1\alpha_1 + 2k_2\alpha_2\big) \end{aligned} \right.
\ee
We then proceed exactly as for the propagator, knowing that for each configuration of signs, the
$\{6j\}$ is peaked on the classical geometry of the tetrahedron. The perturbative expansion with
respect to this flat geometry is thus given by the Gaussian moments of the Hessian matrix
$H_{\{\eps\}}$ of $\Phi_{\{\eps\}}$. Let us stress that, in contrast with the previous studies of
the asymptotics of the 6j-symbol, we are here scaling the lengths of the tetrahedron (or
equivalently $\di{t}$), keeping the length ratios $k_1$ and $k_2$ fixed, instead of scaling $j_t$.

As for the propagator, the four saddle points give the same contribution, and the two sign
configurations of a given saddle point are related by complex conjugation. This can be done in a
quite explicit way. Introduce $\omega$ to be the truncated Taylor expansion of $\Phi_{\{\eps\}}$,
starting at order three onwards, around the saddle point $(\alpha_1,\alpha_2,0)$ with
$\eps_{12}=-\eps_1=-\eps_2=1$. Let $H^{-1}$ be the corresponding inverse of $H_{\{\eps\}}$:
\be
H^{-1} = \f{1}{2} \begin{pmatrix} \f{1}{1-k_1^2}\cot\thet_t & -\f{1}{\sqrt{1-k_1^2}\sqrt{1-k_2^2}\sin\thet_t} & 0 \\
                                  -\f{1}{\sqrt{1-k_1^2}\sqrt{1-k_2^2}\sin\thet_t} & \f{1}{1-k_2^2}\cot\thet_t & 0 \\
                                  0 & 0 & \f{\tan\thet_t}{1-(k_1^2+k_2^2)} \end{pmatrix}
\ee
We also introduce the volume of the tetrahedron, which enters the Gaussian integrals of $H$:
\be \label{tetrahedron volume}
V_t = \f{\di{t}^3}{12}\ k_1 k_2\sqrt{1-(k_1^2+k_2^2)}.
\ee
The expansion of this isosceles 6j-symbol is then (see appendix.\ref{appB} for more details):
\be \label{6j expansion}
\begin{Bmatrix}
j_1 & j_t & j_t \\
j_2 & j_t & j_t
\end{Bmatrix}
=\f{1}{\sqrt{1-k_1^2}\sqrt{1-k_2^2}\sqrt{12\pi V_t}}\ \sum_{p\geq 0}(-1)^p
\Big(\f{C_{2p}}{\di{t}^{2p}}\ \cos\big(S_R+\f{\pi}{4}\big)\ +\ \f{C_{2p+1}}{\di{t}^{2p+1}}\
\sin\big(S_R+\f{\pi}{4}\big)\Big)
\ee
where the coefficients $C_P$, for $P=2p,2p+1$, are given by finite sums:
\be \label{coeff 6j}
C_P = \sum_{n=0}^P \f{(-1)^n}{(2(P+n))! n!}\ \sum_{\vec{\beta}\in\{1,2,3\}^{2(P+n)}}
\pp^{2(P+n)}_{\vec{\beta}}\Big(f\omega^n\Big)_{|(\alpha_1,\alpha_2,0)}\ H^{-1}_{\vec{\beta}}
\ee
Thus, all even orders are in phase with the leading order asymptotics, given by the original
Ponzano-Regge formula in $\cos(S_R+\pi/4)$. This leading order is easily recovered by computing the
coefficient $C_0$, with $f(\alpha_1,\alpha_2,0)=\sqrt{1-k_1^2}\sqrt{1-k_2^2}$:
\be \label{6j LO}
\begin{Bmatrix}
j_1 & j_t & j_t \\
j_2 & j_t & j_t
\end{Bmatrix}
\,\underset{LO}{\sim}\,
\f{1}{\sqrt{12\pi V_t}}\ \cos\Big(S_R+\f{\pi}{4}\Big).
\ee
On the other hand, all odd orders are out of phase (or in quadrature of phase) with this leading
order. If we were scaling the spin $j_t$ instead of the length $\di{t}/2$, the result would not
have had such a simple structure with sines and cosines being mixed up at all orders (but leading).

This asymptotic series formula for the isosceles tetrahedron shows that only the Regge action is
relevant and no other frequency appears in the 6j-symbol. We believe this feature to generalize to
the generic 6j-symbol since its asymptotics can also be extracted using saddle point techniques
\cite{freidellouapre}.

The coefficient of a given order is simply given by the contractions of the derivatives of
$f\omega^n$ with the Gaussian moments. For a given order $P$, the highest derivatives of $\omega$
involved correspond to $n=1$ in \eqref{coeff 6j} and equals $2(P+1)$. For instance, the NLO is
obtained by setting $P=1$. With the notations of the previous section, we have:
\be
C_1 = f_2\ H^{-1}_2 - (f_1\ \omega_3 + f_0\ \omega_4)H^{-1}_4 + \f{f_0}{2}\omega_3\ \omega_3\
H^{-1}_6
\ee
and introducing the reduced volume $v=V_t/\di{t}^3$ :
\be \label{6j NLO}
\begin{Bmatrix}
j_1 & j_t & j_t \\
j_2 & j_t & j_t
\end{Bmatrix}
\,\underset{NLO}{\sim}\,
\f{1}{\sqrt{12\pi V_t}}\ \cos\Big(S_R+\f{\pi}{4}\Big) - \f{\cos^2\thet_t}{\di{t} \sqrt{12\pi V_t}}\f{P_1(k_1,k_2)}{48(12v)^3}\
\sin\Big(S_R+\f{\pi}{4}\Big),
\ee
where $P(k_1,k_2)$ is a symmetric polynomial in $k_1^2$ and $k_2^2$:
\be
P_1(k_1,k_2) = 3(1-k_1^2)^2(1-2k_1^2)+ 3(1-k_2^2)^2(1-2k_2^2)-3+46k_1^2k_2^2 +25k_1^4k_2^4 \\
- 44(k_1^4k_2^2+k_1^2k_2^4) + 10(k_1^6k_2^2+k_1^2k_2^6).
\ee
This polynomial is not simply related to the volume and we haven't succeeded in providing it with a
geometric interpretation. It would nevertheless be very interesting to understand its geometrical
origin in order to interpret physically the higher order corrections to the graviton propagator.

For extremal values of $k_1$, this polynomial simplifies. We get $P_1(0,k)=3(1-k^2)^2(1-2k^2)$ for
$k_1=0$. At the other end at $k_1=1$, we obtain $P_1(1,k)=-4k^4(1-k^2)$ with obvious roots 0 and 1.
Let us point out that $k_{1,2}$ actually never physically reaches these extreme values 0 and 1, but
its bounds depend on the representation $j_t$ (due to the $\SU(2)$ recoupling theory):
$$
\f1{2\di{t}}\le k_e \le 1-\f1{2\di{t}}.
$$
When $k_1$ reaches these extreme values, the coefficients of $P_1$ are polynomials in $1/\di{t}^2$.
\footnotemark
\footnotetext{
For instance, when the edge $e_1$ is at minimal length, $j_1=0$ or $k_1=\f{1}{2\di{t}}$, the
coefficients of $P_1$ read:
$$
P_1(\f{1}{2\di{t}},k_2) = 3\big(1-\f{1}{\di{t}^2} + \f{5}{16\di{t}^4} -
\f{1}{32\di{t}^6}\big)+\big(-12+\f{23}{2\di{t}^2}-\f{11}{4\di{t}^4}+\f{5}{32\di{t}^6}\big)k_2^2 +
\big(15-\f{11}{\di{t}^2}+\f{25}{16\di{t}^4}\big)k_2^4 + \big(-6+\f{5}{2\di{t}^2}\big)k_2^6.
$$
}

\begin{figure}[t] \begin{center}
\includegraphics[width=5cm]{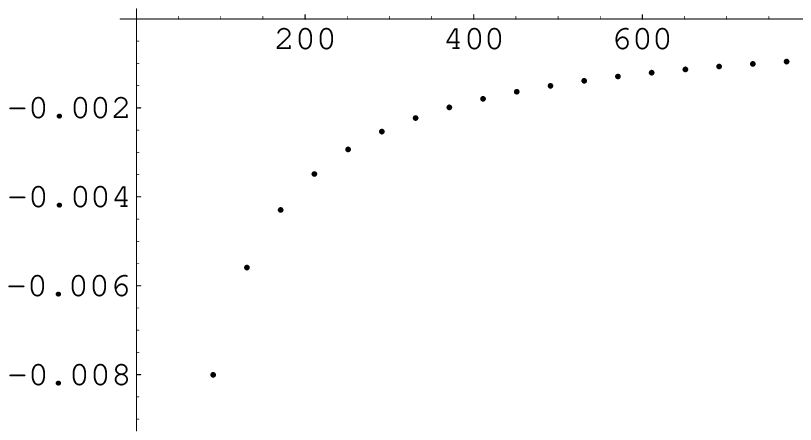}\qquad
\includegraphics[width=5cm]{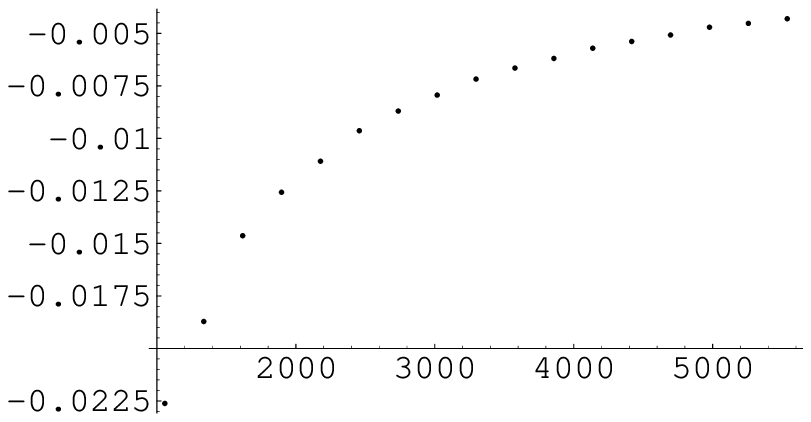}\quad
\includegraphics[width=5cm]{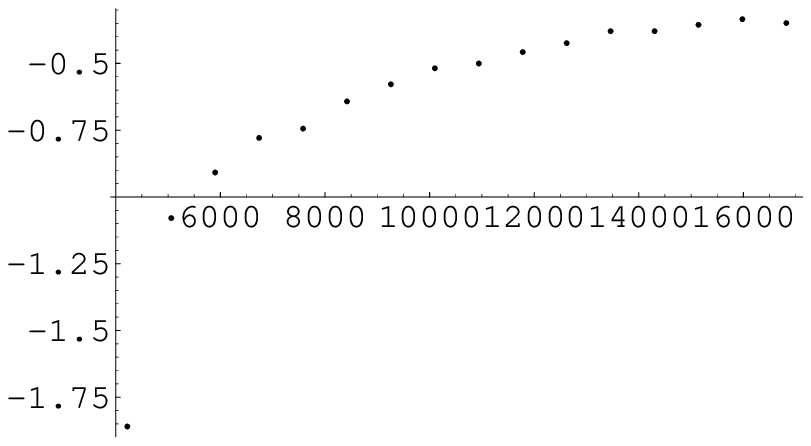}\quad
\caption{ \label{plot 6jNLO} Differences between the 6j-symbol and the analytical result \eqref{6j NLO} for three pairs $(k_1,k_2)=(k,k)$: from left to right, $k=1/2$, $k=3/14$, $k=3/42$. The X axis stands for $\di{t}=N\di{0}$, for $\di{0}$ respectively fixed to 1, 7, and 21, while $N$ goes from 200 to 800.}
\end{center}
\end{figure}

The result \eqref{6j NLO} is confirmed by numerical simulations, see figure \ref{plot 6jNLO}. These
plots represent numerical simulations of the $\{6j\}$ minus the analytical formula \eqref{6j NLO},
for three pairs $(k_1,k_2)$. We have used in these simulations the particular case $k_1=k_2=k$, for
which
\be
P_1(k,k) = (1-k^2)\Big(3-21k^2+55k^4-45k^6\Big)
\ee
whose only root in $[0,1]$ is $k=\f{1}{45}\sqrt{15[(10(81\sqrt{310}+1450))^{\f{1}{3}} +
(10(81\sqrt{310}-1450))^{\f{1}{3}}+55]}\approx 0.8248$ thus inducing the vanishing of the NLO.
To enhance the comparison, we have multiplied by $\di{t}^{5/2}$ to see how the coefficient of the
NLO is approached, and suppressed the oscillations by dividing by those of the NNLO,
$\cos(S_R+\f{\pi}{4})$. The numerics support both the coefficient and the phase.

Notice that in the equilateral situation, $k_1=k_2=1/2$ represent (half) the length ratios but also
(half) the spin ratios. We can thus switch easily to the usual $1/j_t$ expansion:
\begin{align}
\{6j_t\}^{NLO} &= \f{2^{5/4}}{\sqrt{\pi \di{t}^3}}\cos\big(S_R+\f{\pi}{4}\big) - \f{31}{72\cdot 2^{1/4}\sqrt{\pi\di{t}^5}}\sin\big(S_R+\f{\pi}{4}\big) \\
&= \f{\cos\big(S_R+\f{\pi}{4}\big)}{2^{1/4}\sqrt{\pi j_t^3}} - \f{1}{2^{9/4}\sqrt{\pi j_t^5}}\Big[3\cos\big(S_R+\f{\pi}{4}\big) + \f{31\cdot 2^{5/2}}{576}\sin\big(S_R+\f{\pi}{4}\big)\Big]
\end{align}
This point of view shows that it is much more natural to study the asymptotics of the 6j-symbol in
term of the inverse length $1/\di{}$ instead of the inverse spin label $1/j$. For instance, the
leading order coefficient is given in term of the volume $V_t$ of the tetrahedron with edge lengths
given by the $d_j$'s and not the $j$'s.

Finally, we point out that the asymptotics given above in term of the cosine and sine of the Regge
action holds for mid-range values of $k_1,k_2$ and it breaks down for $k_1,k_2$ close to their
extremal values $0$ and $1$. Indeed when $k_2=0$ the asymptotics are given in term of Airy
functions while when $k_2=1$ they are given by the (non-oscillatory) exponential of the Regge
action. The interested reader can find details and references in \cite{physical bound state}.

\section*{Conclusions}

We have shown it is possible to compute analytically the two-point function -- the graviton
propagator -- at all orders in the Planck length for the 3d toy model (the Ponzano-Regge model for
a single isoceles tetrahedron) introduced in \cite{3d toy model}. This builds on the previous work
\cite{corrections to 3d model} where the leading order and first quantum corrections were computed
using the asymptotics of the 6j-symbol in term of the Regge action. Here, we introduced a
representation of the relevant 6j-symbol and of the full graviton propagator as group integrals
over $\SU(2)$. Then one obtains the expansion of the two-point function as a power series in the
inverse spin label (or equivalently in the Planck length) by expanding these group integrals around
their saddle points. We computed explicitly the first and second order corrections to the leading
order behavior and matched them successfully against numerical simulations.

A side-product of these calculations is the corrections to the Ponzano-Regge asymptotic formula for
the 6j-symbol for an isosceles tetrahedron (when four representations are taken equal). We obtain a
series alternating cosines and sines of the Regge action for the tetrahedron (shifted by $\pi/4$).
We computed explicitly the next-to-leading order correction and checked it numerically. An open
issue is the geometrical interpretation of the polynomial coefficient $P_1(k_1,k_2)$ in front of
this first order correction.

To conclude, we have shown how to carry out the calculations of the spinfoam graviton propagator at
all orders at least in this simple setting. We hope to apply the present methods and tools to more
refined 3d triangulations \cite{costas} and to compute spinfoam correlations for 4d quantum gravity
along the lines of \cite{etera simone,grav,Alesci,newvertices,elena}.

\section*{Acknowledgements}

The plots and numerical data presented here were computed using Mathematica 5.0.
MS is grateful for the hospitality of the Perimeter Institute for Theoretical Physics.
Research at Perimeter Institute is supported by the Government of Canada through Industry Canada
and by the Province of Ontario through the Ministry of Research \& Innovation.

\appendix
\section{Details for the propagator expansion} \label{details propagator}


The key object containing the quadratic fluctuations and their corrections is the generating
function $Z(J)$, which is the Gaussian integral of the Hessian matrix $A$ with a source $J$,
evaluated at each saddle point:
\be
Z_{\{\eps\}}(J) = \int dX\ e^{-\f{\di{t}}{2}XA_{\{\eps\}}X+JX},
\ee
with $X=(\phi_1,\phi_2,u)$. The Hessian matrix is given, for all configurations of signs, by:
\be
A_{\{\eps\}} = \begin{pmatrix} 2 \f{1-k_1^2}{\sin\thet_t}\ e^{i\eps_1 \eps_2 \eps_{12}(\f{\pi}{2}-\thet_t)} & 2 \f{k_1 k_2}{\sin\thet_t \cos\thet_t}\ e^{-i\eps_{12}\f{\pi}{2}} & 0 \\
 2 \f{k_1 k_2}{\sin\thet_t \cos\thet_t}\ e^{-i\eps_{12}\f{\pi}{2}} & 2 \f{1-k_2^2}{\sin\thet_t}\ e^{i\eps_1 \eps_2 \eps_{12}(\f{\pi}{2}-\thet_t)} & 0 \\
 0 & 0 & -2i\eps_1 \eps_2 \eps_{12}\ \left[1-(k_1^2+k_2^2)\right] \f{\cos\thet_t}{\sin\thet_t} \end{pmatrix}.
\ee
Due to the initial symmetry between the groups elements $g_1$ and $g_2$, clearly expressed in
\eqref{norm group integral}, $A_{\{\eps\}}$ is invariant under the reversing of $\eps_1$ and
$\eps_2$. $A_{\{\eps\}}$ has also the property of transforming into its complex conjugated matrix
when reverting all signs, an operation which does not change the saddle point
$(\bar{\phi}_1,\bar{\phi}_2,0)$ considered, and equivalently under the flipping of $\eps_{12}$. A
straightforward calculation yields:
\begin{align}
Z_{\{\eps\}}(J) &= Z_{\{\eps\}}\ e^{\f{1}{2\di{t}}JA_{\{\eps\}}^{-1} J} \\
\label{Z}
\text{with} &\quad Z_{\{\eps\}} = \Big(\f{\pi}{\di{t}}\Big)^{3/2}\ \f{-i\eps_1\eps_2\eps_{12}}{(1-k_1^2)(1-k_2^2)\sqrt{2\lvert\cos\thet_t\rvert}}\ e^{\f{i}{2}\eps_1\eps_2\eps_{12}\thet_t}, \\
\text{and} &\quad A^{-1}_{\{\eps\}} = \f{1}{4} \begin{pmatrix} \f{1}{1-k_1^2} & \eps_1\eps_2\f{\cos\thet_t}{k_1 k_2}\ e^{i\eps_1\eps_2\eps_{12}\thet_t} & 0 \\
                                                        \eps_1\eps_2\f{\cos\thet_t}{k_1 k_2}\ e^{i\eps_1\eps_2\eps_{12}\thet_t} & \f{1}{1-k_2^2} & 0 \\
                                                        0 & 0 & \f{2}{1-(k_1^2+k_2^2)}\tan\thet_t\ e^{i\eps_1\eps_2\eps_{12}\f{\pi}{2}} \end{pmatrix}. \label{A(eps)inv}
\end{align}
$A^{-1}_{\{\eps\}}$ and $Z_{\{\eps\}}$ benefit from the previously mentioned symmetries of $A_{\{\eps\}}$. The symmetry flipping $\eps_1$ and $\eps_2$ means that the saddle points $(\alpha_1,\alpha_2,0)$ and $(\pi-\alpha_1,\pi-\alpha_2,0)$ have the same Hessian matrices, for a fixed $\eps_{12}$. Moreover, flipping $\eps_{12}$ while going to the saddle points $(\alpha_1,\pi-\alpha_2,0)$ and $(\pi-\alpha_1,\alpha_2,0)$ does not change $A_{\{\eps\}}$ and $Z_{\{\eps\}}$, up to a change of sign for the non-diagonal coefficients of $A_{\{\eps\}}$.

Let us focus on the normalization $\curly{N}$. The Gaussian moments are generated by successive
derivations of $Z(J)$ with respect to the source, and they are contracted with the derivatives of
$f\exp\di{t}\Omega_{\{\eps\}}$, which we expand into powers of $\di{t}$. We thus have:
\be \label{norm expansion1}
\curly{N} = \f{1}{32\pi^2}\sum_{\eps_1,\eps_2,\eps_{12}} \eps_1\eps_2\ Z_{\{\eps\}}\ e^{\di{t}S_{\{\eps\}}(\bar{\phi}_1,\bar{\phi}_2,0)} \sum_{\substack{N\in\mathbb{N} \\ \vec{\beta}\in\{1,2,3\}^{2N}}} \sum_{n\geq 0} \f{1}{(2N)!\ n!\ \di{t}^{N-n}}\ \pp^{2N}_{\vec{\beta}}\big(f \Omega_{\{\eps\}}^n\big)_{|\substack{\bar{\phi}_1,\bar{\phi}_2,\\ u=0}}\ A^{-1}_{\{\eps\},\vec{\beta}}
\ee
where the correlators $A^{-1}_{\{\eps\},\vec{\beta}}$ are defined according to Wick's theorem:
\be
A^{-1}_{\{\eps\},\vec{\beta}} = \sum_{\substack{\mathrm{all\ possible\ pairings} \\ \mathrm{of}\ (\beta_1,\dots,\beta_{2N})}} A^{-1}_{\{\eps\},\beta_{i_1}\beta_{i_2}}\dots A^{-1}_{\{\eps\},\beta_{i_{2N-1}}\beta_{i_{2N}}}
\ee

As $\Omega$ is a Taylor expansion into powers of $(\phi_1-\bar{\phi}_1)$, $(\phi_2-\bar{\phi}_2)$
and $u$, whose minimal order is 3, the power $n$ of $\Omega$ in \eqref{norm expansion1} is bounded
from above by $N$: $3n\leq 2N$, and the sum over $n$ is thus finite for each $N$. The power of
$1/\di{t}$ receives two contributions: one, positive, from the Gaussian moments, and the other,
negative from the expansion of $\exp\di{t}\Omega_{\{\eps\}}$. We can identify the coefficients of a
given order by the simple change of variables $P=N-n$. Introducing the explicit expressions of
$Z_{\{\eps\}}$ and $S_{\{\eps\}}(\bar{\phi}_1,\bar{\phi}_2,0)$:
\begin{gather} \label{norm expansion2}
\curly{N} = \f{-1}{32(1-k_1^2)(1-k_2^2)\sqrt{2\pi\lvert\cos\thet_t\rvert}\ \di{t}^{3/2}} \sum_{P\in\mathbb{N}} \f{u_P}{\di{t}^P} \\
\text{with}\qquad u_P =  \sum_{n=0}^{2P} \sum_{\vec{\beta}\in\{1,2,3\}^{2(P+n)}} \sum_{\eps_1,\eps_2,\eps_{12}}i\eps_1\eps_2\eps_{12}\ e^{-i\eps_1\eps_2\eps_{12}(2d_{j_t}-\f{1}{2})\thet_t} \f{1}{(2(P+n))!\ n!}\pp^{2(P+n)}_{\vec{\beta}}\big(f\Omega_{\{\eps\}}^n\big)_{|\substack{\bar{\phi}_1,\bar{\phi}_2,\\ u=0}}\ A^{-1}_{\{\eps\},\vec{\beta}}
\end{gather}
Let us further simplify the coefficients $u_P$ by performing the sums over $\eps_1$, $\eps_2$ and
$\eps_{12}$. First, notice that the sign of the imaginary part of $\Omega_{\{\eps\}}$ is
$\eps_1\eps_2\eps_{12}$. Since $f$ is real, and considering the symmetry properties of
$A_{\{\eps\}}^{-1}$ given in \eqref{A(eps)inv}, it is clear that when the signs $\eps_1$, $\eps_2$
and $\eps_{12}$ are all flipped, the derivatives are evaluated at the same saddle point and $u_P$
is transformed into its complex conjugate. Thus, let us work with a fixed value of
$\eps_1\eps_2\eps_{12}$, say 1, and consider the basic properties of the functions $f$ and
$S_{\{\eps\}}$ minus the linear parts in $\phi_1$ and $\phi_2$ (its derivatives greater than three
are those of $\Omega$). More precisely, we are interested in how these functions and their
derivatives, evaluated at a given saddle point, transform when the saddle point is changed. Let us
see for instance the differences when going between the saddle points
$(\bar{\phi}_1=\alpha_1,\bar{\phi}_2=\alpha_2)$ and
$(\bar{\phi}_1=\pi-\alpha_1,\bar{\phi}_2=\alpha_2)$.

Having impose the value of $\eps_1\eps_2\eps_{12}$, this change of saddle point is determined by
the flips of $\eps_1$ and $\eps_{12}$. We have: $f(\pi-\phi_1,\phi_2,u)=f(\phi_1,\phi_2,u)$ and
$(\phi_{12}^+ +\phi_{12}^-)(\pi-\phi_1,\phi_2,u)=2\pi -(\phi_{12}^+
+\phi_{12}^-)(\phi_1,\phi_2,u)$, while the real part of $S_{\{\eps\}}$ is non-zero only when
derivated an even number of times. Thus, $f\Omega^n(\phi_1,\phi_2,u)$ equals
$f\Omega^n(\pi-\phi_1,\phi_2,u)$ when we flip in the same time $\eps_{12}$ in front of
$(\phi_{12}^+ +\phi_{12}^-)$ in $S_{\{\eps\}}$. This means that each derivation with respect to
$\phi_1$ flips the sign between the two saddle points considered. There is now three possibilities:
(i)such a derivation is contracted with another derivation w.r.t. $\phi_1$ through
$A_{11}^{-1}=\f{1}{4(1-k_1^2)}$, then the sign is changed twice, i.e. there is no change of sign.
(ii)It is contracted with a derivation w.r.t. $u$ via $A^{-1}_{1u}$ which is zero, so that there is
in fact no contribution. (iii)It is contracted with a derivation w.r.t. $\phi_2$ via $A^{-1}_{12}$
whose sign changes under the flip of $\eps_1$. Thus these two saddle points give the same
contribution. The proof can be repeated between the four saddle points. Finally:
\be
\curly{N} = \f{-1}{4(1-k_1^2)(1-k_2^2)\sqrt{2\pi\lvert\cos\thet_t\rvert}\ \di{t}^{3/2}} \sum_{P\in\mathbb{N}} \f{N_P}{\di{t}^P}
\ee
with $N_P$ given by \eqref{coeff norm}.

The same analysis can be performed for the numerator of the propagator. One has simply to take into
account the fact that the insertion $\f{k_1
k_2}{4\cos^2\thet_t}(a_{\{\eps\}}+b_{\{\eps\}}/\di{t}+c_{\{\eps\}}/\di{t}^2)$ involves three
different powers of $\di{t}$. To perform the sums over the signs $\eps_1$, $\eps_2$ and
$\eps_{12}$, first notice, like for the denominator, that flipping of all them three transforms the
coefficients into its complex conjugate. Then, restricting attention to a fixed value of the
product $\eps_1\eps_2\eps_{12}$, it is easy to check that the derivative of $a$, $b$ and $c$ w.r.t.
$\phi_1$ evaluated at $\phi_1=\pi-\alpha_1$ is equal to $(-1)^{p_1}$ times that evaluated at
$\phi_1=\alpha_1$, while flipping $\eps_1$ and $\eps_{12}$, with $p_1$ being the number of
derivatives w.r.t. $\phi_1$. The same is true for $\phi_2$. Thus, we can reproduce the previous
argument showing that the four saddle points give the same contribution. This leads us to:
\be
W_{1122} = \f{-k_1 k_2}{16(1-k_1^2)(1-k_2^2)\cos^2\thet_t\sqrt{2\pi\lvert\cos\thet_t\rvert}\ \curly{N}\ \di{t}^{3/2}}\Big\{\f{1}{2\di{t}}\sum_{i,j=1,2}\pp^2_{ij}a\ A^{-1}_{ij} + \sum_{P\geq2} \f{W_P}{\di{t}^P}\Big\}
\ee
with $W_P$ given by \eqref{coeff prop}.

\section{Details for the expansion of the 6j-symbol} \label{appB}

Let us compute the generating function:
\begin{gather}
Z^{(6j)}_{\{\eps\}}(J) = \int dX\ e^{-\f{\di{t}}{2}X\tl{H}_{\{\eps\}}X + JX} \\
\text{with}\quad
\tl{H}_{\{\eps\}} = \f{2i\eps_{12}}{\sin\thet_t} \begin{pmatrix} \eps_1\eps_2 (1-k_1^2)\cos\thet_t & \sqrt{1-k_1^2}\sqrt{1-k_2^2} & 0 \\
 \sqrt{1-k_1^2}\sqrt{1-k_2^2} & \eps_1\eps_2 (1-k_2^2)\cos\thet_t & 0 \\
 0 & 0 & -\eps_1 \eps_2 \left[1-(k_1^2+k_2^2)\right]\cos\thet_t \end{pmatrix}.
\end{gather}
Taking care of the fact that $\tl{H}_{\{\eps\}}$ has purely imaginary coefficients, one has:
\begin{align}
Z^{(6j)}_{\{\eps\}}(J) &= Z^{(6j)}_{\{\eps\}}\ e^{\f{1}{2\di{t}}J\tl{H}_{\{\eps\}}^{-1} J} \\
\label{Z 6j}
\text{with} &\quad Z^{(6j)}_{\{\eps\}} = \f{\pi^2}{\sqrt{1-k_1^2}\sqrt{1-k_2^2}\sqrt{12\pi V_t}}\ e^{-i\eps_1\eps_2\eps_{12}\f{\pi}{4}}, \\
\text{and} &\quad \tl{H}_{\{\eps\}}^{-1} = \f{i\eps_{12}}{2} \begin{pmatrix} \f{\eps_1\eps_2}{1-k_1^2}\cot\thet_t & \f{-1}{\sqrt{1-k_1^2}\sqrt{1-k_2^2}\sin\thet_t} & 0 \\
\f{-1}{\sqrt{1-k_1^2}\sqrt{1-k_2^2}\sin\thet_t} & \f{\eps_1\eps_2}{1-k_1^2}\cot\thet_t & 0 \\
0 & 0 & \f{\eps_1\eps_2}{1-(k_1^2+k_2^2)}\tan\thet_t \end{pmatrix},
\end{align}
where the volume $V_t$ is given by \eqref{tetrahedron volume}.
Using \eqref{regge action}, we obtain an expression similar to \eqref{norm expansion2}:
\be
\begin{Bmatrix}
j_1 & j_t & j_t \\
j_2 & j_t & j_t
\end{Bmatrix} = \f{1}{8\sqrt{1-k_1^2}\sqrt{1-k_2^2}\sqrt{12\pi V_t}} \sum_{P\geq 0} \f{\tl{C}_P}{\di{t}^P}
\ee
with the series coefficients in term of the Hessian:
\be
\tl{C}_P = \sum_{n=0}^{2P} \f{1}{(2(P+n))!\ n!} \sum_{\eps_1,\eps_2,\eps_{12}} (i\eps_{12})^n\ e^{-i\eps_1\eps_2\eps_{12}(S_R+\f{\pi}{4})}
\sum_{\vec{\beta}\in\{1,2,3\}^{2(P+n)}} \ \pp^{2(P+n)}_{\vec{\beta}}\big(f \omega_{\{\eps\}}^n\big)_{|\substack{\bar{\phi}_1,\bar{\phi}_2,\\ u=0}}\
\tl{H}^{-1}_{\{\eps\},\vec{\beta}}
\ee
We are now in position to repeat the arguments of the previous section. The symmetries of the
functions $f$, $i\eps_{12}\omega$, combined with those of $\tl{H}^{-1}_{\{\eps\}}$ imply that the
four saddle points contribute the same. Moreover, the two configurations of signs corresponding to
a given saddle point, which are related by flipping $\eps_1$, $\eps_2$ and $\eps_{12}$, are related
by complex conjugation. The coefficient $\tl{C}_P$ is thus completely determined by the saddle
point $(\alpha_1,\alpha_2,0)$ with $\eps_{12}=-\eps_1=-\eps_2=1$. Writing
$\tl{H}^{-1}_{\{\eps\}}=i\eps_{12} H^{-1}_{\{\eps\}}$, we have that
$\tl{H}^{-1}_{\{\eps\},\vec{\beta}}= (i\eps_{12})^{P+n} H^{-1}_{\{\eps\},\vec{\beta}}$, and:
\be
\tl{C}_P = 8\ \Re\Big(i^{P-2\lfloor \f{P}{2}\rfloor}e^{-i(S_R+\f{\pi}{4})}\Big) \sum_{n=0}^{2P} \f{(-1)^{n+\lfloor \f{P}{2}\rfloor}}{(2(P+n))!\ n!} \sum_{\vec{\beta}\in\{1,2,3\}^{2(P+n)}} \ \pp^{2(P+n)}_{\vec{\beta}}\big(f \omega^n\big)_{|(\alpha_1,\alpha_2,0)}\ H^{-1}_{\vec{\beta}}\ _{|\eps_{12}=-\eps_1=-\eps_2=1}
\ee
It is then clear that $\Re\Big(i^{P-2\lfloor \f{P}{2}\rfloor}e^{-i(S_R+\f{\pi}{4})}\Big)$ is simply $\cos\big(S_R+\f{\pi}{4}\big)$ for even $P$, and $\sin\big(S_R+\f{\pi}{4}\big)$ for odd $P$.


\end{document}